\documentclass[10pt,a4paper,final]{article}

\usepackage[backref,
            natbib
            ,style = numeric-comp
            ,maxnames = 2
            ,backend = bibtex
            ,sorting=none
            ]{biblatex}
  
\usepackage[T1]{fontenc} 
\usepackage[utf8]{inputenc} 
\usepackage{palatino,eulervm}
\usepackage[english]{babel} 
\usepackage{csquotes}
\usepackage{amsmath,amssymb,amsfonts,amsthm} 
\usepackage[table,dvipsnames*,svgnames]{xcolor}
\usepackage{fixltx2e}
\usepackage{showkeys}
\usepackage{nicefrac}
\usepackage{slashed}
\usepackage{tikz}
\usetikzlibrary{arrows}

\usepackage[top=2.5cm, bottom=2.5cm, outer=3cm, inner=3cm, heightrounded, marginparwidth=2cm, marginparsep=0.5cm]{geometry} 
\usepackage[active]{srcltx}
\usepackage{extarrows}

\usepackage{braket}
\usepackage{mathrsfs}
\usepackage{marginnote}
\usepackage{wrapfig}

\makeatletter

\let\@authors\@empty
\let\@email\@empty
\let\@affiliationone\@empty
\let\@affiliationtwo\@empty
\let\@pdfsubject\@empty
\let\@keywords\@empty
\let\@preprint\@empty

\providecommand{\pdfsubject}[1]{\gdef\@pdfsubject{#1}}
\providecommand{\keywords}[1]{\gdef\@keywords{#1}}
\renewcommand{\author}[1]{\ifx\@authors\@empty\toks@\expandafter{#1}\else\toks@\expandafter{\@authors, #1}\fi\edef\@authors{\the\toks@}}
\providecommand{\email}[1]{\ifx\@email\@empty\toks@\expandafter{#1}\else\toks@\expandafter{\@email, #1}\fi\edef\@email{\the\toks@}}
\providecommand{\affiliationone}[1]{\gdef\@affiliationone{#1}}
\providecommand{\affiliationtwo}[1]{\gdef\@affiliationtwo{#1}}
\providecommand{\preprint}[1]{\gdef\@preprint{#1}}

\makeatother

\usepackage[hyperindex,breaklinks]{hyperref}

\hypersetup{
    unicode=false,          
    pdftoolbar=true,        
    pdfmenubar=true,        
    pdffitwindow=false,     
    pdfstartview={FitH},    
    pdfproducer={Brown University}, 
    pdfnewwindow=true,      
    colorlinks=true,       
    linkcolor=black,          
    citecolor=DarkGreen,        
    filecolor=magenta,      
    urlcolor=cyan!30!black!70         
}

\usepackage{graphicx}

\allowdisplaybreaks[3]
\providecommand{\texorpdfstring}[2]{#1}

\newcommand{\del}{\partial}

\newcommand{\bal}{\begin{align}}
\newcommand{\eal}{\end{align}}
\newcommand{\scri}{\mathscr{I}}

\DeclareMathOperator{\tr}{tr}

\renewcommand{\[}{\begin{equation}}
\renewcommand{\]}{\end{equation}}

\newcommand{\wkeq}{\overset{w}{=}}

\bibliography{symmetry.bib}

\begin{document}

\title{Noether's Second Theorem and Ward Identities for Gauge Symmetries}

\author{Steven G.\ Avery\textsuperscript{a}}
 \email{\href{mailto:steven\_avery@brown.edu}{steven\_avery@brown.edu}}

\affiliationone{%
\textsuperscript{a,b}Brown University\\Department of Physics\\182 Hope St, Providence, RI 02912}

\author{Burkhard U.\ W.\ Schwab\textsuperscript{b}}
 \email{\href{mailto:schwab@cmsa.fas.harvard.edu}{schwab@cmsa.fas.harvard.edu}}

\affiliationtwo{%
\textsuperscript{b}Harvard University\\Center for Mathematical Science and Applications\\1 Oxford St, Cambridge, MA 02138}

\keywords{}
\pdfsubject{}
\preprint{Brown-HET-1687}


\makeatletter
\thispagestyle{empty}

\begin{flushright}
\begingroup\ttfamily\@preprint\par\endgroup
\end{flushright}

\begin{centering}
\begingroup\Large\normalfont\bfseries\@title\par\endgroup
\vspace{1cm}

\begingroup\@authors\par\endgroup
\vspace{5mm}

\begingroup\itshape\@affiliationone\par\endgroup
\vspace{3mm}
\begingroup\itshape\@affiliationtwo\par\endgroup
\vspace{3mm}

\begingroup\ttfamily\@email\par\endgroup
\vspace{0.25cm}

\begin{minipage}{14cm}
 \begin{abstract}\normalsize
Recently, a number of new Ward identities for large gauge transformations and large diffeomorphisms have been discovered. Some of the identities are reinterpretations of previously known statements, while some appear to be genuinely new. We use Noether's second theorem with the path integral as a powerful way of generating these kinds of Ward identities. We reintroduce Noether's second theorem and discuss how to work with the physical remnant of gauge symmetry in gauge fixed systems. We illustrate our mechanism in Maxwell theory, Yang--Mills theory, $p$-form field theory, and Einstein--Hilbert gravity. We comment on multiple connections between Noether's second theorem and known results in the recent literature. Our approach suggests a novel point of view with important physical consequences.
 \end{abstract}
\end{minipage}

\vspace{3mm}
\rule{\textwidth}{.5mm}
\vspace{-1cm}

\end{centering}

\makeatother

\tableofcontents
\newpage

\section{Introduction}
\label{sec:intro}

Gauge symmetry sometimes appears to be a curious shell game. One starts with some initial global symmetry algebra and makes it ``local'' via the introduction of new degrees of freedom, enlarging the symmetry algebra enormously; then, states that differ by gauge transformations are identified as the same physical state, effectively reducing the symmetry algebra. It is typically expected that the reduced symmetry algebra relating physical observables is the same as the initial algebra. In which case, the net effect of the gauge procedure, is to introduce new dynamical degrees of freedom (the gauge bosons). In the end, the advantage of the redundant description over a description involving only physical degrees is that the \emph{physical} description is nonlocal. Of course, in the context of gravity, global symmetries lose their meaning, and we are forced to work with gauged symmetries.

It has long been known that for gravity in asymptotically flat space~\cite{Bondi:1962px, Sachs:1962wk} or asymptotically AdS\textsubscript{3}~\cite{Brown:1986nw}, the final physical symmetry algebra is an infinite-dimensional enhancement of the ``global part'' of the gauge group. Only recently, however, has it been realized that the enhancement also occurs for higher dimensional gravity, Maxwell theory,  Yang--Mills theory, and string theory, and moreover, that the symmetry constrains the IR structure via nontrivial Ward identities~\cite{He:2014cra, He:2014laa, He:2015zea, Kapec:2014zla, Kapec:2015ena, Kapec:2015vwa, Lysov:2014csa, Strominger:2013jfa, Strominger:2013lka, Strominger:2014pwa, Strominger:2015bla, Avery:2015gxa}.

This has led to a veritable explosion in activity on the amplitude side, with work on the leading and subleading soft limit in gravity \cite{He:2015yua,Zlotnikov:2014sva,Campiglia:2014yka,Kalousios:2014uva,Broedel:2014bza}, Yang--Mills theory \cite{Broedel:2014fsa,Bern:2014oka,Bern:2014vva,Klose:2015xoa}, (ambitwistor) string theory \cite{DiVecchia:2015oba,DiVecchia:2015bfa,Bianchi:2015yta,Geyer:2014lca,Schwab:2014sla,Schwab:2014fia,Bianchi:2014gla,Lipstein:2015rxa}, supersymmetric theories \cite{Rao:2014zaa,Liu:2014vva}, and theories in higher dimensions \cite{Schwab:2014xua,Geyer:2014lca,Adamo:2014yya}.

The present article provides a general path integral formalism for writing Ward identities for these ``large'' gauge symmetries. Our starting point is Noether's second theorem, which constrains the general structure of theories with local symmetry. This allows us to relate the Ward identities to two-form charges used in the above literature. This approach seems considerably more general and powerful than other approaches when a path integral formulation is available, allowing one to quickly write down Ward identities for new theories and symmetries. Beyond the technical advantages we exhibit, our approach requires a significantly different point of view with a number of physical consequences, which we detail below.

The Ward identities we write down are \emph{not} for ``large'' gauge transformations, but rather for \emph{residual} gauge transformations \emph{after} imposing a gauge condition. It is worth emphasizing that if one uses the Ward identities we write without fixing the gauge symmetry, one can derive a number of nonsensical conclusions. This is not surprising, since the path integral is ill-defined until one gauge fixes. While, as we argue below, residual gauge symmetry is necessarily large, the converse is not true. In our formalism, this explains the reduction of $BMS^+\times BMS^-$ to the diagonal $BMS^0$ in~\cite{Strominger:2013jfa}: one must solve the residual diffeomorphism equation \emph{everywhere} and propagate the boundary conditions from $\scri^-$ to $\scri^+$. 

Residual gauge symmetries are symmetries of the gauge fixed action, but they may not be symmetries of the initial and final wavefunctionals that determine the initial and final field configuration. In the language of the path integral, when one starts and ends in the vacuum, this effect is synonymous with spontaneous symmetry breaking. Since gauge transformations are in general inhomogeneous shifts of the gauge field, the currents associated with these transformations are necessarily linear in the fields. This leads to an interpretation of the residual gauge symmetry as inserting physical states into correlators, which we understand as Goldstone modes. This is true in the case of the Abelian gauge field as well as in linearized theories of Yang--Mills and gravity. The soft charges of \cite{He:2014laa,He:2015zea,Kapec:2015ena,Strominger:2013lka,Strominger:2013jfa} fall into this category; they insert soft states since they are defined at asymptotic infinity. The familiar case of AdS\textsubscript{3} is different, since AdS behaves like an IR regulator. But this notion is broader and includes the proposal of \cite{Gaiotto:2014kfa} to interpret the full photon field as encoding two Goldstone modes since residual gauge symmetries contain transformations that give finite charges when integrated over any finite hypersurface in the bulk.

Thinking about residual gauge symmetry thus takes the focus away from the boundary of the spacetime manifold $\mathcal{M}$. Since one can consider a path integral for subregions $R\subset M$ (causal diamonds are particularly natural since they provide a unitary ``sub-theory''), one can write identities on these for subregions. We suspect this may be relevant to recent speculations in~\cite{Hawking:2015qqa}. Regardless, insofar as the soft photon theorem -- or its cousin, electromagnetic memory~\cite{Susskind:2015hpa} -- is relevant for terrestrial experiments, it is desirable to have an understanding of the Ward identity that does not depend on the causal structure of the entire universe, which is not even asymptotically flat!\footnote{See~\cite{Andrade:2015fna} for a potentially different approach to this issue.}

Let us note that our approach does come with a drawback: while physically it is clear that different gauge conditions must yield equivalent results, mathematically it is frequently unclear that this is true. We suspect that a fuller understanding of this issue may be realized in the Batalin--Vilkovisky (BV) formalism.

The current paper is organized as follows. In Sec.~\ref{sec:noeth-second-theor}, we review Noether's first and second theorems for classical field theory. In Sec.~\ref{sec:ward-identities}, we apply these results to the path integral and derive Ward identities. In Sec.~\ref{sec:examples}, we demonstrate our approach with a number of illustrative examples. We conclude in Sec.~\ref{sec:discussion}.

\section{Noether's theorems}
\label{sec:noeth-second-theor}

Noether's 1918 theorem~\cite{Noether1918}\footnote{See~\cite{Noether:1918zz} for an English translation available on the arXiv.} relating infinitesimal ``global'' symmetries to conservation laws, is a cherished cornerstone of modern theoretical physics; however, the second theorem (appearing in the same work) applicable to ``local'' symmetry remains somewhat obscure~\cite{Kosmann}.\footnote{This assertion is based, in part, on informal discussions. An important exception is~\cite{Barnich:2001jy}, which introduced the authors to Noether's second theorem. It also appears to be (somewhat) known in the BV quantization and mathematical literature; see e.g.~\cite{Gomis:1994he, Fulp:2002fm} and~\cite{Olver}, respectively.} Our goal is to use Noether's \emph{second theorem} as a starting point for a general approach to Ward identities for gauge symmetry. In particular, we are motivated by recent new Ward identities for large gauge symmetry in gravity and QED~\cite{He:2014cra, He:2014laa, He:2015zea, Kapec:2015ena, Kapec:2015vwa, Lysov:2014csa, Strominger:2013jfa, Strominger:2013lka, Strominger:2014pwa}, and recent discussions in~\cite{Gaiotto:2014kfa}.

It is well-known that Noether's theorem, charge conservation, and the symplectic structure have a special character when there is ``local'' symmetry; and, of course, symmetries of gauge theory correlators is an old subject. Indeed many or even most of the statements appearing here appear in some form in the literature; however, we hope to present them in a novel, streamlined form that is useful for recent and future developments. Of particular note are the seminal works~\cite{Lee:1990nz, Iyer:1994ys, Barnich:2001jy}, which we draw heavily from. See also~\cite{Abbott:1977ug, Abbott:1982jh, Bak:1993us, Silva:1998ii, Julia:1998ys, Julia:2000er, Julia:2002df}.

\subsection{Notation and terminology}

When discussing Noether's theorems it is important to distinguish between identities that hold only after applying the equations of motion, and identities that hold universally. Throughout our discussion, we use $=$ to denote equality without using equations of motion and call this a \emph{strong} or \emph{off-shell} equality. We use $\wkeq$ to denote equality only after using equations of motion and call this a \emph{weak} or \emph{on-shell} equality. This is the language of Dirac~\cite{Dirac, Dirac:1950pj}; but note that we use $\wkeq$ instead of the more generic $\approx$.

Following common usage, we use \emph{local symmetry} to denote symmetry transformations that are parametrized by functions of spacetime, and \emph{global symmetry} to denote symmetry transformations that are at most part of a countable set. We use \emph{gauge} to indicate transformations that do not affect physical observables. Of course, local symmetry must be \emph{gauged}, but the converse is not true. As a slight abuse, we call local symmetry transformations with bounded support \emph{small gauge transformations} and those with support on the boundary of the theory \emph{large gauge transformations}. In this language, an important point is that large gauge transformations need not be gauged. Whether or not large gauge transformations are gauged should be determined by physical considerations.

When discussing in generality we use $\phi(x)$ to implicitly denote the entire field content, $\phi^a(x)$, where the index $a$ could label different scalar fields as well as components of vector or tensor fields, etc. We hope the reader can fill in implicit summations without difficulty.

We may also switch between index notation and the language of differential forms, as expedient. All appearing forms follow the standard convention 
\[\omega^{(p)} = \frac{1}{p!}\omega_{\mu_{1}\cdots\omega_{\mu_{p}}}dx^{\mu_{1}}\wedge\cdots\wedge dx^{\mu_{p}}\] and the Hodge dualizer $\star$ is taken to act only on the closest form in a wedge product
\[\star \omega\wedge\eta = (\star\omega)\wedge \eta\] to minimize the amount of brackets needed. 

\subsection{Noether's first}
\label{sec:noether-first}

Consider a particular infinitesimal transformation $\hat{\delta}\phi$ for which the action functional is invariant up to a possible boundary term:
\begin{equation}
S[\phi + \hat{\delta}\phi] = S[\phi] + \int d^d x\, \partial_\mu K^\mu,
\end{equation}
without imposing equations of motion. Noether's first theorem constructs a current whose divergence is proportional to the equations of motion, and therefore is conserved on-shell. There are two equivalent derivations.

The first begins by noting that a general transformation of the action takes the form
\begin{equation}
\delta S = \int d^{d}x\big(-E(\phi) \delta\phi + \partial_\mu \theta^\mu(\phi; \delta\phi)\big),
\end{equation}
where $E(\phi)$ are the equations of motion (ie. Euler--Lagrange derivatives),\footnote{We find it convenient to define $E$ with an extra minus sign from~\cite{Lee:1990nz}.} and $\theta$ is the ``symplectic potential current density'' in~\cite{Lee:1990nz}. If we use the symmetry transformation $\hat{\delta}\phi$, then we find the conservation law
\begin{equation}
\partial_\mu j^\mu = E(\phi)\hat{\delta}\phi \wkeq 0 \qquad j^\mu = \theta^\mu(\phi; \hat{\delta}\phi) - K^\mu.
\end{equation}
This is essentially the original approach in~\cite{Noether1918}. 

The second is to deform the symmetry with an arbitrary function $\rho(x)$ as
\begin{equation}
\delta_\rho\phi(x) = \rho(x)\hat{\delta}\phi(x).
\end{equation}
Then, on the one hand, locality of the transformation and the fact that $\delta_\rho$ is a symmetry for constant $\rho$ implies\footnote{We assume the action only has explicit dependence on $\phi$ and its first derivative; otherwise there could be terms with higher derivatives of $\rho$. (This is the case for the Einstein--Hilbert action.) Note, however, that when we restrict to $\rho$ with compact support, one can integrate by parts ``for free'' and put $\delta_\rho S$ into this form.}
\begin{equation}\label{eq:deformed-deltaS}
\delta_\rho S = \int d^dx\, \big(\theta^\mu\partial_\mu\rho + \rho \,\partial_\mu K^\mu\big);
\end{equation}
while on the other hand, if $\rho(x)$ has compact support there can be no boundary term, and thus
\begin{equation}\label{eq:j-def}
\delta_\rho S = - \int d^dx\, \rho\partial_\mu j^\mu \wkeq 0.
\end{equation}
We take this as the \emph{definition} of the Noether current $j$, since this is what naturally enters into the Ward identity. Finally, varying $\rho(x)$ with compact support tells us that 
\begin{equation}
\partial_\mu j^\mu \wkeq 0.
\end{equation}

\subsection{Noether's second}
\label{sec:noether-second}

Noether's second theorem applies when one has a collection of infinitesimal symmetries $\delta_\lambda\phi$ parametrized by one or more arbitrary functions $\lambda(x)$, i.e., local symmetry. Note that the first theorem continues to hold for local symmetry (even ``small'' gauge transformations). Noether's second theorem gives us strong identities, which constrain the form of the equations of motion and the current $j^\mu$. Again there are two approaches. 

For simplicity, we focus on the case when the transformation may be written in the form\footnote{We stray from the literature in using $f$ instead of $R$ here.}
\begin{equation}\label{eq:f-delta}
\delta_\lambda\phi = f(\phi)\, \lambda + f^\mu(\phi)\partial_\mu \lambda,
\end{equation}
but it is straightforward to consider transformations, as Noether did, involving arbitrarily high derivatives of $\lambda$. (Although, the authors know of no physically interesting examples.) Let us start with
\begin{equation}
\delta_\lambda S = \int d^{d}x\big(-E \delta_\lambda\phi + \partial_\mu \theta^\mu(\phi; \delta_\lambda\phi)\big)
\end{equation}
and note that the contribution of the boundary term $\theta$ must vanish if $\lambda$ has compact support, and the left-hand side vanishes because this is a symmetry.\footnote{Potential boundary terms cannot contribute for $\lambda$ with compact support, by locality.} Thus,
\begin{equation}
\int d^{d}x \,E(\phi) \delta_\lambda\phi = 0\qquad \text{($\lambda$ with compact support)}.
\end{equation}
Now, we may vary this equation with respect to $\lambda$ with compact support, and we find the strong identity, using the notation in~\cite{Barnich:2001jy}
\begin{equation}
\Delta(E) = 0\qquad \Delta(\cdot) = f(\phi)(\cdot) - \partial_\mu\big(f^\mu(\phi)\cdot\big),
\end{equation}
where $\Delta$ is the adjoint differential operator of $\delta_\lambda$ in~\eqref{eq:f-delta}, in the Sturm--Liouville theory sense. This is essentially the approach taken in~\cite{Noether1918}.\footnote{Note that there are two more independent Noether identities one may find (for local symmetry transformations with up to one derivative); however, the content is entirely contained in the final result in~\eqref{eq:gauge-j}: one gives the two-form $k$, and the other relates the current $j$ to $S$ and $k$ (for constant $\lambda$).} This means the equations of motion are not all independent, and therefore the Cauchy problem is not well-posed. Since we demand that physical observables are uniquely determined, even in the classical theory and even without going to the Hamiltonian formulation, we see that some degrees of freedom are gauge.

As emphasized in~\cite{Barnich:2001jy}, this statement has important implications. Since the operator $\Delta$ was defined via integration by parts, we have
\begin{equation}\label{eq:def-Smu}
E\delta_\lambda\phi = \lambda(x)\Delta\big(E \big) + \partial_\mu S^\mu(E ; \lambda) = \partial_\mu S^\mu(E; \lambda).
\end{equation}
One sees that the current $S^\mu$ (defined by integrating by parts) satisfies the same equation that the canonical Noether current $j^\mu$ satisfies, with the important difference that $S^\mu$ vanishes on-shell. Since the whole current vanishes, any conserved charges one might define as $\int_\Sigma \star S_\lambda$ on some spacelike surface $\Sigma$ identically vanish on-shell.

On the other hand,  $j^\mu$ and $S^\mu$ must (assuming a trivial de Rham cohomology) differ by the divergence of a two-form
\begin{equation}
\partial_\mu (j^\mu(\lambda) - S^\mu(\lambda)) = 0 \qquad \Longrightarrow \qquad j^\mu(\lambda) = S^\mu(\lambda) + \partial_\nu k^{\nu\mu}(\lambda),\qquad
k^{\mu\nu}(\lambda) = - k^{\nu\mu}(\lambda).
\end{equation}
Thus, the charge (defined for $j$) must be given by a codimension-2 integral computing the flux of $k$ through $\sigma = \partial \Sigma$, and the only nonvanishing charges are those for which $\lambda$ is nonzero on $\sigma$.

Let us note that classically speaking, favoring $j^\mu$ over other possible ``Noether currents'' that satisfy $\partial_\mu j^\mu = E\delta\phi$ might seem ad hoc; however, when considering the path integral, the current defined by~\eqref{eq:j-def} is singled out because it appears in the Ward identity. Similarly, when discussing the classical theory, one may worry about the ambiguities introduced by adding boundary terms to the action that do not change the equations of motion; however, in the path integral, boundary terms can be absorbed into initial and final states, $\Psi_{0,f}$ below.

\begin{figure}[htb]
  \centering
\includegraphics{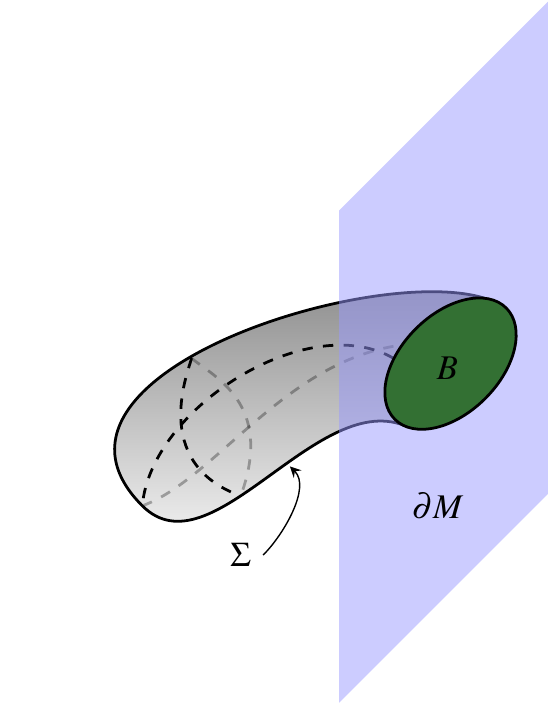}
  \caption{A depiction of the indicator $\mathbf{1}_R(x)$ for $R$ having support on the boundary of the spacetime manifold $M$. The support of $R$ on $\partial M$ is denoted $B = R \cap \partial M = \partial R \cap \partial M$. The interior portion of $\partial R$ is denoted $\Sigma$.}
  \label{fig:indicator}
\end{figure}

The second approach is to take $\rho(x)$ in the previous section to be an ``indicator'' function for some subregion $R$ of the spacetime $M$:
\begin{equation}\label{eq:indicator}
\rho(x) = \mathbf{1}_R(x) = 
\begin{cases}
1 & x \in R\\
0 & x \notin R
\end{cases}.
\end{equation}
Let $R$ be a simply connected compact region. Then~\eqref{eq:deformed-deltaS} takes the form
\begin{equation}\label{eq:compact-R}
\delta_R S = \int_R d^d x\,\partial_\mu(K^\mu - \theta^\mu) = -\int_{\partial R} d^{d-1}x_\mu j^\mu \wkeq 0.
\end{equation}
Now, formally break $\partial R$ into two disjoint regions: $\partial R = \Sigma_1 \cup \Sigma_2$. It follows that
\begin{equation}
\int_{\Sigma_1 }d^{d-1}x_\mu j^\mu(\lambda) \wkeq - \int_{\Sigma_2}d^{d-1}x_\mu j^\mu(\lambda),
\end{equation}
for all $\lambda$. Consider a $\lambda$ that vanishes on $\partial \Sigma_1 = \partial \Sigma_2 = \sigma$. For every such $\lambda$, we can define $\lambda'$ such that $\lambda'\to \lambda$ in the neighborhood of $\Sigma_1$ and $\lambda' \to 0$ in the neighborhood of $\Sigma_2$. (A similar argument was used in~\cite{Lee:1990nz} to show that the charge for any small gauge transformation must vanish.) Thus, it follows from locality that
\begin{equation}
\int_{\Sigma_1}d^{d-1}x_\mu j^\mu(\lambda) = \int_{\Sigma_1}d^{d-1}x_\mu j^\mu(\lambda') \wkeq
-\int_{\Sigma_2}d^{d-1}x_\mu j^\mu(\lambda') = 0.
\end{equation}
To wit, the flux of $j$ through any codimension one surface vanishes on-shell if $\lambda$ vanishes on the boundary of the surface:
\begin{equation}
\int_{\Sigma}d^{d-1}x_\mu j^\mu(\lambda) \wkeq 0\quad \text{for all $\lambda\to 0$ on $\partial\Sigma$}.
\end{equation}
It follows that (at least locally)
\begin{equation}
j^\mu(\lambda) \wkeq \partial_\nu k^{\nu\mu}(\lambda),
\end{equation}
and therefore
\begin{equation}\label{eq:gauge-j}
j^\mu(\lambda) = S^\mu(\lambda) + \partial_\nu k^{\nu\mu}(\lambda)\qquad S^\mu(\lambda) \wkeq 0.
\end{equation}
For our purposes, this is the characteristic feature of local symmetry.

\subsection{Two-form currents}
\label{sec:two-form-currents}

Unlike most of the relevant literature, the focus of our discussion here is not on defining conserved charges; however, note that the Noether charge for local symmetry is the integral of the two-form $k$ over a (codimension-two) sphere at infinity. The two-form $k$ and charges are discussed in e.g.~\cite{Lee:1990nz, Barnich:2001jy, Iyer:1994ys, Bak:1993us, Julia:1998ys, Julia:2000er, Julia:2002df, Silva:1998ii}, and general higher-form symmetries in~\cite{Gaiotto:2014kfa, Sharpe:2015mja}. The fact that in a gauge theory charge is given by the flux of a two-form on a codimension-two surface at infinity sounds like Gauss's law for electromagnetism; however, there are two important generalizations: first, we have an arbitrary function $\lambda$ on $\sigma$, and second this argument holds for any local symmetry without discussing the form of the equations of motion.

For our purposes, the critical property of local symmetry is the fact that the Noether current can be written in terms of a two-form $k$ as in~\eqref{eq:gauge-j}. For $\lambda$ constant, one recovers the conserved current that couples to the gauge field. If one can find $\lambda$, ``asymptotic reducibility parameters'' in~\cite{Barnich:2001jy}, for which the $j^\mu(\lambda)$ flux weakly vanishes on the boundary of spacetime and which respect the boundary conditions of the theory, then one may use $k$ to define asymptotically conserved charges. Generally, one expects to recover the constant $\lambda$ symmetry algebra in this way, but sometimes one finds enhanced symmetry as occurs quite famously in asymptotic $\text{AdS}_3$~\cite{Brown:1986nw} and in asymptotically flat space~\cite{Bondi:1962px, Sachs:1962wk}.

In gravity, the two-form $k$ in~\eqref{eq:gauge-j} gives the ADM and Bondi mass, and black hole entropy~\cite{Wald:1993nt, Iyer:1994ys}. In the literature, there is much discussion of ambiguities in the definition of the charge. In particular, the addition of a boundary term to the action does not change the equations of motion, but shifts $\theta$. Moreover, if one defines the current via $\partial_\mu j^\mu = E\delta\phi$, then the two-form is ill-defined, and one could even use $S^\mu$. As alluded to above, these ambiguities are not an issue for Ward identities: boundary terms play a physical role in the path integral and are not arbitrary, and the current that enters into the Ward identity is unique. One could still argue that $k$ is only defined up to the addition of the exterior derivative of the Hodge star of a one-form; however, we always integrate $k$ over a compact surface, so the ambiguity never contributes in any calculation.

\section{Implications for QFT}
\label{sec:ward-identities}

Having reviewed the classical consequences of local symmetry, let us now turn to the consequences for correlators in quantum field theory. Symmetries of the classical theory become Ward identities for correlators in quantum field theory. As in the classical case, there are a couple of twists when considering local symmetry. Previous discussions of the two-form $k$ focus on defining charges, which with a symplectic structure, translate into statements in an operator language for QFT. We focus on using the path integral to directly make statements about correlators, bypassing some of the complications inherent to the operator approach.

\subsection{Gauge fixing}

In order to correctly define the path integral, we need to eliminate the enormous overcounting of gauge equivalent configurations. As should be clear from classical considerations, small gauge transformations (for which $\lambda \to 0$ on $\partial M$) must describe physically equivalent points in phase space. Following the Fadeev--Popov procedure, one imposes a gauge condition on $\phi$ in such a way as to (ideally) slice through all gauge orbits once.

An important point, at this time, is to define ``good'' gauge fixing conditions. A good gauge fixing condition should eliminate all of the local degrees of freedom, but keep representatives of the large gauge transformations unfixed. For example, in Lorenz gauge, the residual gauge parameters satisfy the Laplace equation  $\Box \lambda = 0$. The residual gauge condition becomes a well-posed boundary value problem, such that $\lambda$ in the interior of a region is uniquely determined by its value on the boundary. This is the critical property of a good gauge condition:
\begin{equation}\label{eq:G-lambda}
\lambda(x) = \int_{\partial M} d^{d-1}y\, G(\phi; x, y) \lambda(y),
\end{equation}
with Green's function $G(\phi; x, y)$ becoming a delta function $\delta(y-y_0)$ as $x\in M$ approaches a point $y_0$ on $\partial M$. (We are focusing on infinitesimal gauge transformations that are connected to the identity.) We put in $\phi$ in the Green's function, because for interacting non-Abelian theories the gauge fixing condition frequently depends on the background. This means we have a field dependent gauge transformation, and $\lambda$ should be thought of as an operator in Ward identities. In particular, it cannot pull out of the path integral, among other restrictions.

In addition to the above, there are additional constraints on the space of allowed large gauge transformations from demanding that boundary conditions are preserved on $\partial M$. We shall discuss these as they arise in specific applications. 

Let us emphasize that the Ward identities we write down are for the residual gauge symmetry which is a symmetry of the gauge fixed action; there are no additional contributions from the gauge fixing term or ghost sector. This is a slightly different point of view from much of the literature. A necessary condition for a gauge symmetry to be residual is that it be large in the usual sense, but it is not sufficient. This is how we explain the reduction from $BMS^+\times BMS^-$ to $BMS^0$ in~\cite{Strominger:2013jfa}: one must solve the residual diffeomorphism equation \emph{everywhere} and propagate the boundary conditions from $\scri^-$ to $\scri^+$.

In situations in which one uses more than one coordinate or gauge patch (as in most discussions of BMS, in cases of nontrival topology, magnetic monopoles, etc), then one must carefully match the gauge fixing conditions in the overlap. Different gauge patches will have different Green's functions $G$, but one should match the value of $\lambda$ on the interface between regions. This is all consistent with the idea that these Ward identities may be written for subregions, either by slicing the path integral open or by considering the ``sub-theory'' with boundary source terms. Basically, one should imagine pasting different Ward identities together to get the identity for the entire theory.

\subsection{Ward identities}

We treat the path integral with initial and final wave function(al)s, $\Psi_0$ and $\Psi_f$, along with local insertions, $\Phi_j$:
\begin{equation}
\Braket{\Phi_1(x_1)\cdots \Phi_n(x_n)}_{\Psi_f, \Psi_0} = \int D\phi\,\Psi_f^*(\phi_f)\Psi(\phi_0)\Phi_1(x_1)\cdots \Phi_n(x_n)e^{i S[\phi]},
\end{equation}
where $\phi_{0,f}$ are the values of $\phi$ on the ``initial'' and ``final'' boundary of $M$. We denote these boundaries as $\Sigma_0$ and $\Sigma_f$, respectively. For asymptotically flat spacetimes these will be $\scri^-\cup i^-$ and $\scri^+\cup i^+$.\footnote{One has several choices on how to treat spatial infinity, $i^0$, none of which seem to affect the physical conclusions for reasonable assumptions.} We include $\Psi_f$ and $\Psi_0$ explicitly for two reasons: first, as we discuss below large gauge symmetries are typically spontaneously broken so that even in the vacuum the transformation of the boundaries contributes; and second, as we briefly revisit in the conclusion, we would like to be able to apply our results to causal diamonds inside a larger spacetime.

\begin{figure}[htpb]
  \centering
\includegraphics{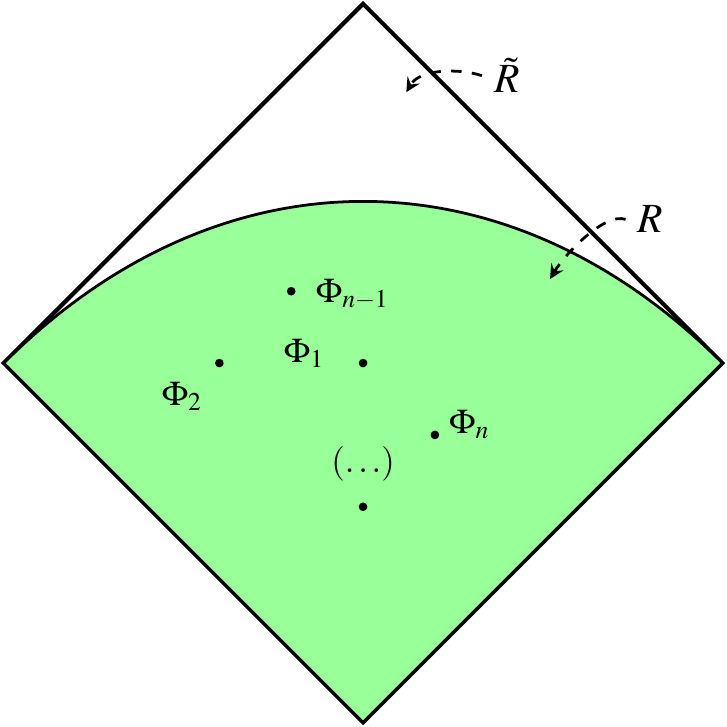}
  \caption{A depiction of the region $R$ and its complement $\tilde{R}$ as used to find the global Ward identity~\eqref{eq:global-ward-id}. $R$ encloses all interior insertions and the entire past boundary $\Sigma_0$.}
  \label{fig:R-Rc}
\end{figure}

Consider some local, infinitesimal transformation of the field $\phi$: $\delta\phi$. The measure $D\phi$ is invariant under shifts (ignoring the possibility of anomalies, for now), and thus performing a change of variables in the path integral yields the Schwinger--Dyson equation,
\begin{equation}
\int D\phi\, \left[\delta\big(\Psi_f^*(\phi_f)\Psi(\phi_0)\Phi_1(x_1)\cdots \Phi_n(x_n)\big) + i\,\delta S\big(\Psi_f^*(\phi_f)\Psi(\phi_0)\Phi_1(x_1)\cdots \Phi_n(x_n)\big)\right]e^{iS(\phi)} = 0,
\end{equation}
or 
\begin{equation}\begin{aligned}\label{eq:S-Deq}
0 &= \delta\Braket{\Phi_1(x_1)\cdots \Phi_n(x_n)}_{\Psi_f, \Psi_0}\\
   &= \Braket{\delta\big(\Phi_1(x_1)\cdots \Phi_n(x_n)\big)}_{\Psi_f, \Psi_0}
  + \Braket{\Phi_1(x_1)\cdots \Phi_n(x_n)}_{\delta\Psi_f, \Psi_0} 
  + \Braket{\Phi_1(x_1)\cdots \Phi_n(x_n)}_{\Psi_f, \delta\Psi_0}\\
  &\hspace{136pt}+ i\Braket{\delta S\, \big(\Phi_1(x_1)\cdots \Phi_n(x_n)\big)}_{\Psi_f, \Psi_0}.
\end{aligned}\end{equation}
One gets additional anomaly terms if the measure is not invariant. Let us use the deformed symmetry transformation $\delta_\rho$ with $\rho$ given by the indicator function~\eqref{eq:indicator} for some subregion $R$. Rewriting~\eqref{eq:deformed-deltaS} as
\begin{equation}
\delta_\rho S = \int_M d^dx\,\big(j^\mu\partial_\mu\rho + \partial_\mu(\rho K^\mu)\big),
\end{equation}
we see with $\rho$ the indicator function, we have
\begin{equation}
\delta_\rho S = \int_{\partial M \cap R} d^{d-1}x_\mu\big(j^\mu + K^\mu\big) - \int_{\partial R} d^{d-1}x_\mu j^\mu.
\end{equation}
Noting that $\partial M\cap R = \partial M \cap \partial R$, and breaking $\partial R = \Sigma \cup B$ with $B$ the boundary part of $R$, we get
\begin{equation}
\delta_\rho S = \int_B d^{d-1}x_\mu K^\mu - \int_\Sigma d^{d-1}x_\mu j^\mu.
\end{equation}
See Fig.~\ref{fig:indicator} for a depiction of the surfaces. When $R$ is a compact region, one gets~\eqref{eq:compact-R} as a special case. All of this applies equally well to global or to local symmetry transformations. In the case of local symmetry, however, we know that $j^\mu$ takes the form~\eqref{eq:gauge-j}, in which case
\begin{equation}
\delta_\rho S = \int_B d^{d-1}x_\mu K^\mu - \int_\Sigma d^{d-1}x_\mu S^\mu - \int_{\partial \Sigma}d^{d-2}x_{\mu\nu} k^{\mu\nu},
\end{equation}
with $K$, $S$, and $k$ all implicitly depending on $\lambda$.

If at this point one blindly writes down Ward identities using~\eqref{eq:S-Deq} without gauge fixing, then one comes to a number of embarrassing conclusions, including that propagators for gauge fields must identically vanish, cf.~\cite{Gomis:1994he}. This is an artifact of using the improper path integral, which integrates over all gauge redundancy. The solution is of course to correctly define the path integral via Fadeev--Popov and an appropriate gauge fixing condition. Then, we may write down Ward identities for the residual gauge symmetry. (More general identities can also be written down related to BRST symmetry; however, we focus on the residual gauge symmetry here.) By definition, these are also symmetries of the gauge fixing term and the Fadeev--Popov ghosts.

Let us now consider the simplest class of Ward identities: when the region $R$ does not extend to the boundary of $M$. In this case, we find
\begin{equation}
\delta_\rho S = - \int_{\Sigma} d^{d-1}x_\mu\, S^\mu(\lambda),
\end{equation}
which recall weakly vanishes, so we should expect a somewhat trivial identity. For no insertions in the path integral, we find
\begin{equation}
\Braket{\int_{\Sigma} d^{d-1}x_\mu\, S^\mu(\lambda)}_{\Psi_f, \Psi_0} = 0.
\end{equation}
Also, if $x_1, \dots, x_n$ are outside of the region $R$:
\begin{equation}
\Braket{\Phi_1(x_1)\cdots \Phi_n(x_n)\int_{\Sigma} d^{d-1}x_\mu\, S^\mu(\lambda)}_{\Psi_f, \Psi_0} = 0.
\end{equation}
On the other hand, consider the case where we have a single insertion inside the surface $\Sigma = \partial R$, then
\begin{equation}
i\Braket{\Phi_1(x_1)\int_{\Sigma} d^{d-1}x_\mu\, S^\mu(\lambda)}_{\Psi_f, \Psi_0} 
  = \Braket{\delta_\lambda\Phi_1(x_1)}_{\Psi_f, \Psi_0};
\end{equation}
$\Phi_1$ transforms under the gauge transformation. Note that this identity would be rather shocking if we had not already discussed the gauge fixing condition which determines $\lambda$ in the neighborhood of $x_1$ in terms of its value on $\Sigma$. This identity can be derived by using~\eqref{eq:def-Smu} and the standard Schwinger--Dyson equation: 
\begin{equation}
i\Braket{\Phi_1(x_1)\int_{\Sigma} d^{d-1}x_\mu\, S^\mu(\lambda)}_{\Psi_f, \Psi_0}
= i\Braket{\Phi_1(x_1)\int_{R} d^dx_\mu\, E\big(\phi(x)\big)\delta_\lambda\phi(x)}_{\Psi_f, \Psi_0}
= \Braket{\delta_\lambda\Phi_1(x_1)}_{\Psi_f, \Psi_0}
\end{equation}
The identities with more insertions inside and outside of $R$, follow in the obvious way.

Now, let us consider the more interesting case when $R$ extends to the boundary of $M$. In this case, the two-form may make a contribution. Note that from~\eqref{eq:G-lambda} it follows that $\lambda$ vanishes everywhere unless $\lambda$ has some support on $\partial M$. Let us start with no insertions in the path integral, then one finds
\begin{equation}
\int_B d^{d-1}x_\mu \Braket{K^\mu}_{\Psi_f, \Psi_0} - \int_{\partial \Sigma}d^{d-2}x_{\mu\nu}\Braket{k^{\mu\nu}}_{\Psi_f, \Psi_0} = -i\delta_{R, \lambda}\Braket{\Psi_f|\Psi_0},
\end{equation}
Since there are no insertions inside $\Sigma$, we dropped the $S^\mu$ term. For most gauge theories, the gauge symmetry is a symmetry of the Lagrangian without an additional boundary term, and therefore $K^\mu = 0$. This is not true for Chern--Simons theories, however. Basically, in the simplest case we find that the insertion of the integral of $k$ is equivalent to shifting the boundary conditions on the path integral.

We see that the three pieces of the Ward identity each play distinct roles: $S^\mu$ transforms the interior insertions in the usual way; $k^{\mu\nu}$ shifts the boundary conditions of the path integral; and $K^\mu$ is a source/nonconservation term from the noninvariance of the Lagrangian. Note that this last piece can be treated as an additional shift of $\Psi_{0, f}$.

\subsection{Global identities}
\label{sec:global-id}

It is frequently convenient to use a trick with the Ward identity, to immediately get the ``global'' form of the Ward identity.

Consider the path integral with $n$ interior insertions as above, with $\ket{\Psi_0} = \ket{\Psi_f} = \ket{0}$ where $\ket{0}$ is some fiducial vacuum state. (We are explicitly considering the case of spontaneously broken symmetry, cf.~\cite{Matsumoto:1973hg, Matsumoto:1974nt, Nakanishi:1974pz}.) Then choose $R$ to encompass all of the interior insertions and the entire past boundary, as shown in the Fig~\ref{fig:R-Rc}. Then, the Ward identity reads
\begin{equation}
\Braket{\delta_\lambda(\Phi_1\cdots \Phi_n)}_{0, 0} 
+ \Braket{\Phi_1\cdots\Phi_n}_{0, \delta_\lambda 0} = 
i\Braket{\delta_{R,\lambda}S\, (\Phi_1 \cdots \Phi_n)}.
\end{equation}
On the other hand, consider the Ward identity for $\tilde{R}$, the complement of $R$:
\begin{equation}\label{eq:R-bar}
\Braket{\Phi_1\cdots\Phi_n}_{\delta_\lambda 0, 0} 
  = i\Braket{\delta_{\tilde{R}, \lambda} S(\Phi_1\cdots \Phi_n)}.
\end{equation}
Then note that $\delta_{\tilde{R}, \lambda} S = -\delta_{R, \lambda}S$ because the normal is oriented in the opposite sense. Thus one arrives at the global version of the Ward identity for spontaneously broken symmetry:
\begin{equation}\label{eq:global-ward-id}
\Braket{\delta_\lambda(\Phi_1\cdots \Phi_n)}_{0, 0} 
+ \Braket{\Phi_1\cdots\Phi_n}_{0, \delta_\lambda 0}
+ \Braket{\Phi_1\cdots\Phi_n}_{\delta_\lambda 0, 0} = 0.
\end{equation}
This identity was recently discussed in this context in~\cite{Avery:2015gxa}.

To see roughly what this says, note that
\begin{equation}
\delta_\lambda\Psi_{0}[\phi_0] = \int_{\Sigma_0} d^{d-1}x \frac{\delta\Psi_{0}[\phi_0]}{\delta\phi_0(x)}\delta_\lambda\phi_0(x).
\end{equation}
For instance for Maxwell theory, let the wavefunctional for $\ket{0}$ be Gaussian, as is the case for the free vacuum. Then, formally we see
\begin{equation}
\delta_\lambda \Psi_{0}[A] \simeq (\text{const.})\,\Psi_{0}[A] \int_{\Sigma_0} d^{d-1}x_\nu \,(F^{\mu\nu}\partial_\mu \lambda),
\end{equation}
where $\Sigma_0$ is the surface on which $\Psi_0$ is defined. We see that the effect of shifting the boundary is to insert a photon.

Obviously it is rather awkward to be working explicitly with $\Psi_0$ and $\Psi_f$; when demonstrating the connection between the Ward identity and the soft theorem below, we take a different approach, more precise and closer to that of~\cite{He:2014cra, He:2014laa, He:2015zea, Kapec:2015ena, Kapec:2015vwa, Lysov:2014csa, Strominger:2013jfa, Strominger:2013lka, Strominger:2014pwa}. Instead, choose $R$ to enclose all interior insertions and not initial and final boundaries where $\Psi_0$ and $\Psi_f$ are defined. The Ward identity, in terms of $j$, reads
\begin{equation}\label{eq:ward-for-soft}
\Braket{\delta_\lambda(\Phi_1\cdots \Phi_n)}_{0, 0} 
  = i\Braket{\left(\int_{\Sigma_2}\star j - \int_{\Sigma_1}\star j\right)\Phi_1\cdots \Phi_n}_{0, 0}.
\end{equation}
Now we can push $\Sigma_2$ arbitrarily close to $\Sigma_f$ and $\Sigma_1$ arbitrarily close to $\Sigma_0$. In the final step, one reinterprets the integral of $j$ on the boundary as the insertion of a (soft) particle. See the explicit examples below.

\subsection{Commutators}
\label{sec:commutators}

\begin{figure}[htpb]
  \centering
\includegraphics{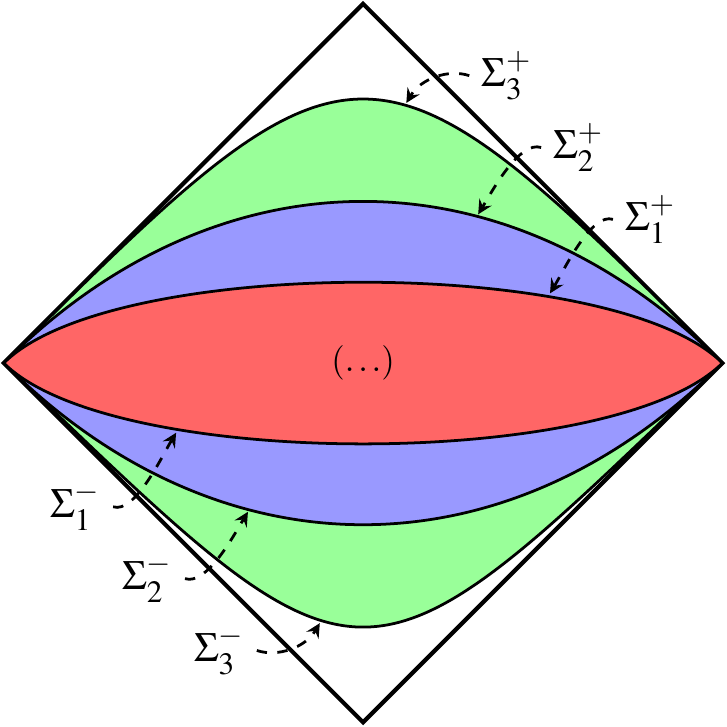}
  \caption{A depiction of the boundaries of $R_1 \subset R_2 \subset R_3$ as used to compute commutators using the Ward identity. It is useful to consider three surfaces to show~\eqref{eq:second-comm-id}, which relates the commutator to one symmetry transformation of the other charge.}
  \label{fig:commutator}
\end{figure}

We can compute commutators using the Ward identity in a way that should be reminiscent of computations in radially quantized two-dimensional CFT. Define three regions $R_{1,2,3}$ such that
\begin{equation}
R_1 \subset R_2 \subset R_3,
\end{equation}
and all three regions have spacelike boundary surfaces $\Sigma_{1,2,3}^{\pm}$, as pictured in Fig.~\ref{fig:commutator}. The details are not important, so long as the boundaries of the regions keep the same time-ordering. We want to imagine that the three boundaries are close together, so that they all share the same enclosed insertions.

A single application of the Ward identity tells us that
\begin{equation}
i\Braket{\int_{\Sigma_1^+-\Sigma_1^-}\star j_{\lambda_1}(\dots)} = \Braket{\delta_{\lambda_1}(\dots)},
\end{equation}
where $(\dots)$ denotes some insertions inside $R_1$. Applying the Ward identity a second time for $R_2$ and gauge parameter $\lambda_2$, one arrives at
\begin{equation}
-\Braket{\int_{\Sigma_2^+ - \Sigma_2^-}\star j_{\lambda_2}\int_{\Sigma_1^+-\Sigma_1^-}\star j_{\lambda_1}(\dots)} = \Braket{\delta_{\lambda_2}\delta_{\lambda_1}(\dots)}.
\end{equation}
To get the other order, one can use $R_3$ and $R_2$ as
\begin{equation}
-\Braket{\int_{\Sigma_3^+ - \Sigma_3^-}\star j_{\lambda_1}\int_{\Sigma_2^+ - \Sigma_2^-}\star j_{\lambda_2}(\dots)} = \Braket{\delta_{\lambda_1}\delta_{\lambda_2}(\dots)}.
\end{equation}
We can write this as two separate identities involving the commutator. First,
\begin{equation}\label{eq:first-comm-id}
\Braket{\int_{\Sigma_2^+ - \Sigma_2^-}\star j_{\lambda_2}\int_{\Sigma_1^+-\Sigma_1^-}\star j_{\lambda_1}(\dots)} 
-\Braket{\int_{\Sigma_3^+ - \Sigma_3^-}\star j_{\lambda_1}\int_{\Sigma_2^+ -\Sigma_2^-}\star j_{\lambda_2}(\dots)}
=\Braket{[\delta_{\lambda_1},\,\delta_{\lambda_2}](\dots)}.
\end{equation}
This can be used to understand the nontrivial consecutive double soft limit of Yang--Mills~\cite{He:2015zea}. The second identity is
\begin{multline}\label{eq:second-comm-id}
\Braket{\int_{\Sigma_2^+ - \Sigma_2^-}\star j_{\lambda_2}\int_{\Sigma_1^+-\Sigma_1^-}\star j_{\lambda_1}(\dots)} 
-\Braket{\int_{\Sigma_3^+ - \Sigma_3^-}\star j_{\lambda_1}\int_{\Sigma_2^+ -\Sigma_2^-}\star j_{\lambda_2}(\dots)} \\
=
-\Braket{\delta_{\lambda_1}\left(\int_{\Sigma^+ -\Sigma^-}\star j_{\lambda_2}\right)(\dots)}
= \Braket{\delta_{\lambda_2}\left(\int_{\Sigma^+ -\Sigma^-}\star j_{\lambda_1}\right)(\dots)},
\end{multline}
where the last line follows from our freedom to slide the surfaces up and down as long as one does not pass over other insertions.

\subsection{Central terms}

As should be familiar from Brown--Henneaux~\cite{Brown:1986nw}, the algebra of large gauge transformations may develop central terms. These terms are not one-loop effects, but are evident at the semiclassical level. They arise from $\delta_{\lambda_1}\int\star k(\lambda_2)$~\cite{Barnich:2001jy}. See~\cite{Terashima:2000gb, Terashima:2001gn} for a path integral derivation of the central term for asymptotically AdS\textsubscript{3} gravity. We rephrase the argument in our language for the general case.

The basic issue is that the bracket on the boundary does not match with the bracket of residual gauge transformations in the interior. From~\eqref{eq:first-comm-id} and~\eqref{eq:second-comm-id}, we have
\begin{equation}
\Braket{[\delta_{\lambda_1},\,\delta_{\lambda_2}](\dots)} = \Braket{\delta_{\lambda_2}\left(\int_{\Sigma^+-\Sigma^-}\star j_{\lambda_1}\right)(\dots)}.
\end{equation}
One would like to rewrite the right-hand side using something like
\begin{equation}\label{eq:no-central-q}
\delta_{\lambda_2}\int_{\Sigma^+-\Sigma^-}\star j_{\lambda_1} \overset{?}{=} \int\star j_{[\lambda_1,\, \lambda_2]},
\end{equation}
where we define the bracket via the commutator
\begin{equation}
[\delta_{\lambda_1},\, \delta_{\lambda_2}]\phi = \delta_{[\lambda_1, \lambda_2]}\phi.
\end{equation}
In general it is not possible to write~\eqref{eq:no-central-q} in the path integral: for $\lambda_1$ and $\lambda_2$ satisfying the residual gauge condition, the bracket $[\lambda_1,\, \lambda_2]$ need not; however, there is a residual gauge transformation taking the same value as $[\lambda_1,\, \lambda_2]$ on the boundary, by using~\eqref{eq:G-lambda}. Let us call that residual gauge transformation $\overline{[\lambda_1,\,\lambda_2]}$. Note that the (classical) charge is the same for $[\lambda_1,\, \lambda_2]$ and $\overline{[\lambda_1,\,\lambda_2]}$, since the charge only depends on the boundary value of $\lambda$. Thus, the central charge is given by
\begin{equation}
  K_{\lambda_1,\lambda_2} = \delta_{\lambda_2}\left(\int\star j_{\lambda_1}\right) - \int\star j_{\overline{[\lambda_1, \, \lambda_2]}}.
\end{equation}

\section{Examples}
\label{sec:examples}

In this section we demonstrate the puissance of Noether's second theorem in conjunction with the path integral formalism to derive a couple of results in gauge theory and gravity.
 
\subsection{Abelian gauge field}
\label{sec:spin-1-gauge}

Let us begin by investigating Maxwell theory. A general gauge transformation for the $U(1)$ vector field $A_{\mu}$ is given by \[A_{\mu}(x)\to A_{\mu}(x) + \del_{\mu}\lambda(x)\] and the action is 
\[S_{1} = \int d^{d}x\ \mathcal{L}_{1} = -\frac{1}{4} \int d^{d}x\ F_{\mu\nu}F^{\mu\nu}.\] The equations of motion are of course $E^{\nu} = \del_{\mu}F^{\mu\nu} = (\star\ d \star F)^{\nu}$. Now, use the statement of Noether's second theorem to derive a current and a two-form from the action. As explained before, variational arguments yield \[\label{eq:noesec} E^{\mu}(A)\delta_{\lambda}A_{\mu} = \lambda(x)\del_{\mu}E^{\mu} + \del_{\mu}S^{\mu}(E^{\mu},\lambda)\] with $\del_{\mu}E^{\mu} = \del_{\mu}\del_{\nu}F^{\mu\nu} = (\star\ d^{2}\star F)= 0$ and $S^{\mu} = \lambda\del_{\nu}F^{\nu\mu} = \lambda E^{\mu}$. By localizing the gauge transformation using a function $\rho(x)$ we can find a weakly conserved current for any gauge transformation
\[j^{\mu} = \del_{\nu}\lambda\, F^{\nu\mu} = [\star (d\lambda \wedge \star F)]^{\mu}.\] As shown in the general treatment, this current, together with the weakly vanishing current $S^{\mu}$ allows us to define the two-form \[k_{\lambda} = \lambda F =\frac12 k_{\lambda,\mu\nu}dx^{\mu}\wedge dx^{\nu}  = \frac12\lambda F_{\mu\nu} dx^{\mu}\wedge dx^{\nu}\label{eq:maxtwoform}\] which is the universal two-form associated with gauge transformations of the $U(1)$ gauge field.

Adding an arbitrary Lagrangian $\mathcal{L}_{\rm matter} = \mathcal{L}(D_{\mu}\Phi^{I},\Phi^{I})$ where $\Phi^{I}$ minimally couples to the photon in a gauge invariant way\footnote{$I$ is an index that enumerates the set of matter fields, which can contain real and complex scalars as well as fermions. $\Phi^{I}$ may have arbitrary charge under the global part of the gauge symmetry.} will not actually alter $k_{\lambda}$ since the possible contributions cancel out when subtracting the weakly vanishing current $S^{\mu}$ from the weakly conserved current $j^{\mu}$. This can be shown rather easily as follows. We take a general variation of the matter Lagrangian $\mathcal{L}(D_{\mu}\Phi^{I},\Phi^{I})$. Then
\[\delta \mathcal{L}(D_{\mu}\Phi^{I},\Phi^{I}) = \delta A_{\mu}\frac{\delta}{\delta A_{\mu}}\mathcal{L}(D_{\mu}\Phi^{I},\Phi^{I}) + \delta \Phi^{J}\frac{\delta}{\delta\Phi^{J}}\mathcal{L}(D_{\mu}\Phi^{I},\Phi^{I})\] where we have variational derivatives on the right-hand side. The second term on the right hand side encodes the equations of motion $E^{J}$ of the matter fields. These take part in Noether's second theorem and do not contribute to the two-form $k_{\lambda}$. The first term is part of the equation of motion of the photon field and corresponds to the set of electric currents in the theory $J_{\Phi^{I}}$. These currents appear in the weakly conserved current $j$ and the weakly vanishing current $S$ with the same sign. We conclude that they cannot contribute to the two-form $k_{\lambda}$. The two-form therefore takes the same form as in the free Maxwell theory. To summarize: while $j$ and $S$ vary, $k_{\lambda}$ is unchanged by minimally coupled matter.

As we have discussed in Sec.~\ref{sec:noether-second}, the charges for small gauge transformations must vanish; however, it is still possible to associate a charge for large gauge transformations, if the manifold on which the theory is formulated has a boundary, and large gauge transformations do exist. If both conditions are fulfilled we choose the indicator function $\rho$ to take nonzero values on the boundary. A charge is given by \[Q(\lambda) = \frac12\int_{\sigma}d^{d-2}x^{\mu\nu}\lambda F_{\mu\nu}\label{eq:abeltwoform}\] where $\sigma=R\cap\del M$, a codimension two surface of the space time, respectively a codimension one surface on the boundary. In the case of flat Minkowski space, we have a choice between letting $\sigma \subset i^{0}$ and $\sigma\subset \scri$. Ref.~\cite{He:2014cra} have shown that for (massless) QED, there is an infinite set of additional charges when looking at $\scri$. These charges follow from choosing $\sigma=\scri^{+}_{-}$ where $\lambda$ can approach finite values, using Stokes' theorem to integrate over all of $\scri$, and the currents of Sec.~\ref{sec:noether-second}. From the gauge invariance of the field strength $F$ follows somewhat unsurprisingly that the algebra of charges is Abelian
\[[Q(\lambda_{1}),Q(\lambda_{2})] = \delta_{1}Q_{2} = -\delta_{2}Q_{1} = 0.\] This implies that the consecutive double soft limit of photons is independent of the order in which they are taken.

In theories with matter, we still use the two-form \eqref{eq:abeltwoform}, but using Stokes' theorem now requires us to restore the electric currents 
\[J^{\mu} = \frac{\delta}{\delta A_{\mu}}\mathcal{L}(D_{\mu}\Phi^{I},\Phi^{I}),\] 
to make $S^{\mu}$ weakly vanishing. The charge is then calculated via 
\[
\int_{\sigma}\star k \wkeq \int_{\Sigma}\star j = \int_{\Sigma} d^{d-1}x_{\mu} (\del_{\nu}\lambda F^{\nu\mu} + \lambda J^{\mu});
\] 
compare with \cite{Kapec:2015ena}.

\paragraph{The soft theorem}
We wish to connect the Ward identity for residual gauge symmetry to the soft theorem, as first established in~\cite{Strominger:2013lka}. This should hopefully highlight the advantages of the path integral approach developed here. Start with the Ward identity as written in~\eqref{eq:ward-for-soft}
\begin{equation}
\Braket{\left(\int_{\Sigma_2}\star j - \int_{\Sigma_1}\star j\right)\Phi_1\cdots \Phi_n} = -i\Braket{\delta_\lambda(\Phi_1\cdots\Phi_n)},
\end{equation}
with initial and final states some particular vacuum and the insertions $\Phi_1$, \dots, $\Phi_n$ having charges $q_1$, \dots, $q_n$. Let us push $\Sigma_2$ up against $\scri^+$ and $\Sigma_1$ against $\scri^-$, then since the two surfaces are after and before all insertions, they can be moved out of the time ordering to act directly on the vacuum:
\begin{equation}
\Braket{\left(\int_{\Sigma_2}\star j - \int_{\Sigma_1}\star j\right)\Phi_1\cdots \Phi_n} =
\left(\bra{0}\int_{\Sigma_2}\star j\right)\, T(\Phi_1\cdots \Phi_n)\ket{0} 
- \bra{0}T(\Phi_1\cdots \Phi_n)\,\left(\int_{\Sigma_1}\star j\ket{0}\right).
\end{equation}
In general the current $j$ takes the form
\begin{equation}
j^\mu = F^{\nu\mu}\partial_\nu\lambda + \lambda J^\mu, 
\end{equation}
where $J^\mu$ is the matter current that sources Maxwell's equation $d\star F = \star J$. By assumption, the vacuum does not have any charged matter, so $\bra{0}J^\mu\ket{0} = 0$. Let us look at the action of the integral insert on the vacuum. Following~\cite{He:2014cra,Kapec:2015ena}, we work in coordinates
\begin{equation}
ds^2 = -dv^2 +2 dvdr + r^2 d\Omega^2,
\end{equation}
with gauge condition
\begin{equation}
A_r = 0,
\end{equation}
and boundary condition
\begin{equation}
A_v(r\to\infty) = 0.
\end{equation}
The residual gauge freedom satisfying the boundary condition is given by arbitrary functions on the sphere $\lambda = \lambda(\theta)$. Then,
\begin{equation}
\int_{\Sigma_1}\star j\ket{0} =
\int_{-\infty}^\infty dv\int d\Omega\, r^2\left(F^{A r}\partial_A \lambda + \lambda J^r\right)\ket{0}.
\end{equation}
The vacuum has photon zero modes, but should not have matter degeneracies, thus
\begin{equation}
\int dv\ J^r \ket{0} = 0.
\end{equation}
This leaves us with the first term, matching with the heuristic argument in Sec.~\ref{sec:global-id}. Since there is no $v$ dependence, this should be a soft photon. Note that asymptotically
\begin{equation}
F^{Ar} = - \frac{1}{r^2}\gamma^{AB}(F_{vB} + F_{rB}) 
  = -\frac{1}{r^2}\gamma^{AB}\big(\partial_vA_B + O(\tfrac{1}{r})\big).
\end{equation}
Then,
\begin{equation}
\int_{\Sigma_1}\star j\ket{0} = -\int_{-\infty}^\infty dv\int d\Omega\, \gamma^{AB}(\partial_v A_B)(\partial_A\lambda)\ket{0}.
\end{equation}
Plugging this and the corresponding result for $\Sigma_2$, and using LSZ reduction for the insertions $\Phi_1$, \dots, $\Phi_n$ reproduces Equation~(7.6) of~\cite{He:2014cra}, in which \citeauthor{He:2014cra} show that it is equivalent to the soft photon theorem. In particular, one should regulate the above integral by putting in a soft Fourier mode $e^{i\omega v}$, and then take the limit as $\omega$ goes to zero. Let us note that the antipodal identification in~\cite{He:2014cra} just results from ensuring that one is using the same residual gauge transformation on $\scri^-$ and on $\scri^+$. It is worth emphasizing that the surface $\Sigma_1$ necessarily cuts across past timelike infinity, where massive particles originate; thus, our result immediately applies to both massless and massive charged matter.

\paragraph{Shift symmetries and large gauge symmetries}
\label{sec:shift-symm-large}

The reader may allow us a quick digression before continuing. When studying low-energy effective Lagrangians, a very important tool for making general statements about the existence of light or massless modes with weak low-energy interactions is Goldstone's theorem. Whenever a global symmetry is spontaneously broken, that is, when it is a symmetry of the action but not of the vacuum, a gapless particle will appear. In relativistic theories, this translates to the masslessness of the Goldstone particle. This particle is characterized by an inhomogeneous transformation---a \emph{shift symmetry}\footnote{The name derives from their action as a translation in field space or equivalently, an infinitesimal shift of the vacuum expectation value of the progenitor field. Note that in the case of a scalar field derived from spontaneous symmetry breaking, the shift $a$ is periodic with period $2\pi$.}---of the field, e.g.,  \[\phi \to \phi + a\] in the easiest case of a Goldstone scalar. Since shift symmetries are global symmetries of the effective action, Noether's first theorem applies. The associated current $j^{\mu}$ is linear in the field at leading order. A corollary of the existence of a shift symmetry is that the effective Lagrangian may only depend on the Goldstone particle via derivatives  $\del_{\mu}\phi$ of the field.\footnote{In light of the subsequent discussion, we may understand $\del_{\mu}\phi$, in a loose sense, as the \emph{field strength} associated with the field $\phi$.} Additionally, the field decouples entirely from the theory in the soft limit, a statement usually known as ``Adler's zero''.\footnote{Essentially because any interaction is proportional to some power of the momentum of the particle.}  Nonlinear corrections in the fields may appear at higher orders in the coupling constant.

An elementary example is the massless scalar. Remember that the action \[S_{0} = \int d^{d}x\ \mathcal{L}_{0} = -\frac12 \int d^{d}x\ \del_{\mu}\phi\del^{\mu}\phi\] is invariant under $\phi(x) \to \phi(x) + a$. The corresponding current is \[j^{\mu}_{a} = a \del^{\mu}\phi\] which is obviously weakly conserved since \[\del_{\mu}j^{\mu}_{a} = a E_{\phi}\wkeq 0\] in the notation used in Sec.~\ref{sec:noeth-second-theor}. Using canonical quantization for $\phi(x)$ we see that the shift symmetry connects the vacuum $|0\rangle$ with a one particle state $|1\rangle$, i.e., \[\langle0|j^{\mu}_{\lambda}|1\rangle = a \langle0|\del^{\mu}\phi(x)|1\rangle = a p^{\mu}e^{ip.x}\] as required by Goldstone's theorem and the matrix element vanishes in the strict soft limit. Current conservation $\langle 0|\del_{\mu}j^{\mu}_{\lambda}|1\rangle = 0$ holds.

It has long been known that Abelian gauge symmetry implies the masslessness of the associated particles. Thus, the photon field is gapless and we want to interpret gauge symmetry as a shift symmetry of the gauge field. The current which connects the vacuum $|0\rangle$ with the one particle state is known \cite{kovner1991photon,Gaiotto:2014kfa} to be the field strength $F_{\mu\nu}$ which satisfies\footnote{This seems distinct from the idea that photons and gravitons are Goldstone bosons of spontaneously broken Lorentz invariance, see e.g., \cite{kraus2002photons}.} \[\langle 0 | F_{\mu\nu}(x) |\epsilon,p\rangle = (p_{\mu}\epsilon_{\nu}-p_{\nu}\epsilon_{\mu})e^{i p.x};\label{eq:abelgoldstone}\] clearly, in the limit $p\to 0$, this matrix element vanishes. The statement of current conservation is the equation of motion $d\star F \wkeq 0$ which is trivially weakly conserved and current conservation holds also in the expectation value. Obviously, we would like to connect this current with the two-form \eqref{eq:maxtwoform}. In clear contrast with the massless scalar, where the shift symmetry was a global symmetry, here the shift symmetry is part of the gauge symmetry, as 
\[A_{\mu}\to A_{\mu} + \del_{\mu}\lambda\] such that $\del_{\mu}\lambda = (\text{const.})$  
More precisely, it has to be part of the \emph{residual part} of gauge symmetry. We are assuming Lorenz gauge here; however, while the constraint equation on the residual gauge varies depending on gauge, the content of the residual gauge symmetry should be unaffected by the choice of gauge. The residual gauge freedom are those functions $\lambda$ which satisfy $\Box \lambda(x) = 0$, i.e., harmonic functions. These are the only functions $\lambda$ which generate symmetries of the fully gauge fixed action with ghost action but aren't (necessarily) symmetries of the vacuum. All the asymptotic symmetries which generate the soft theorems fall into this category. We want to emphasize again that for a complete discussion of this topic, it is absolutely necessary to look at the full gauge fixed action with ghost action.

The two-form $k^{\mu\nu} = \lambda F^{\mu\nu}$ as a \emph{one-form current} in the language of \cite{Gaiotto:2014kfa} satisfies exactly equation \eqref{eq:abelgoldstone} when $\lambda$ is removed from both sides of the equation. In the case of the free Abelian field, this leads one to the conclusion that the photon satisfies all conditions of Goldstone's theorem.\footnote{One may remark that the S-matrix of the free photon is the identity, which means that the photon trivially ``decouples''.} When coupling the theory to some charged matter, the two-form survives since gauge symmetry ought to be conserved. The associated current $j$ gets modified by the electric currents of the charged matter. If we interpret the shift symmetry $\delta A = (\text{const.})$ as part of the residual gauge symmetry, every gauge symmetric Lagrangian retains the shift symmetry of the free Abelian gauge field. As a consequence, we find that the photon continues to behave like a Goldstone mode even when coupling it to matter and stays, importantly, massless. 

\subsection{Yang--Mills theory}
\label{sec:yang-mills}

Consider now Yang--Mills theory with gauge group $G$. It is an easy exercise to use Noether's second theorem on theories with non-Abelian gauge symmetries. One finds essentially the same two-form current as in the Abelian case except that now we also need to sum over adjoint indices. The derivation follows through using the gauge transformation \[\delta_{\lambda}A = d^{\nabla}\lambda\] with the gauge covariant derivative on forms \[d^{\nabla} = d + A\] and the non-Abelian gauge parameter $\lambda = \lambda^{a}T^{a}$. Since the Lagrangian is now \[\mathcal{L}_{1}^{YM} = - \frac12 \tr \star F \wedge F,\] where $F=(d^{\nabla})^{2}$ it is easy to see that $ \star S = \tr \lambda d^{\nabla} \star F$ and $ \star  j =\tr (d^{\nabla}\lambda \wedge \star F)$ such that the two-form $k$ is just \[k = \tr \lambda F\] using partial integration and the trace rule $\tr (A[B,C]) = \tr ([A,B]C)$. 
To define nontrivial charges for the case of residual (large) gauge transformations \cite{Strominger:2013lka,He:2015zea} we may integrate over a codimension two surface $\sigma\subset\scri$ to define a charge for a gauge transformation with parameter $\lambda$ \[Q(\lambda) = \int_{\sigma}\star\tr\lambda F.\]
Notice that, unlike in Abelian gauge theory, the YM field strength $F$ is not gauge invariant but transforms as \[F \to [\lambda,F]\] under infinitesimal gauge transformations. This leads to nontrivial commutators of two charges. We may demonstrate this in the linearized theory over some background $\overline{A}$.  There $A = \overline{A} + a$, the covariant derivative $d^{\nabla} = d^{\overline\nabla} + a = d + \overline{A} + a$, and $\delta_{\lambda}a = d^{\nabla}\lambda$. With these, the two-form is given by
\[k = \tr \left(\lambda d^{\overline\nabla}a\right).\]  Thus, the commutator of $Q(\lambda_{1})$ and $Q(\lambda_{2})$---see Sec.~\ref{sec:commutators}---is given by
\[[Q(\lambda_{1}),Q(\lambda_{2})] = \delta_{1}Q(\lambda_{2}) = Q([\lambda_{1},\lambda_{2}]) + \int_{\sigma}[\lambda_{1},\lambda_{2}]\star\overline{F}\]
where we recognize the last term is independent of the dynamical field $a$. It therefore is a central charge $K_{\lambda_{1},\lambda_{2}}$ \cite{Barnich:2001jy} that depends on the background field configuration at $\scri$. Under usual, physical assumptions, $\overline F$ should be $0$ at null infinity, and we don't expect any nontrivial central charge.

It is possible to derive the soft theorems for gluons from the existence of large (residual) gauge symmetries in this way. Since the calculation is essentially analogous to the one presented above for $U(1)$, we refrain from doing so here. We may also use this result to relate the single soft limit to the commutator of the consecutive double soft limit, which has been investigated in \cite{He:2015zea,Klose:2015xoa} and shown to be nontrivial whenever two gluons of the opposite helicity are taken to be soft in 4D. Using~\eqref{eq:first-comm-id} with the appropriate current of linearized $YM$ theory, $\star j = d^{\overline{\nabla}}\lambda\wedge d^{\overline\nabla}a$, one quickly relates the commutator of the soft limits on the left hand side to the commutator $[\delta_{\lambda_1},\,\delta_{\lambda_2}]$. Since the theory is non-Abelian this commutator is nontrivial.

As a last comment, note that a conclusion about gluons as Goldstone modes similar to the case of Maxwell theory is \emph{not} possible in the full non-linear theory. In particular, since $F$ is now gauge covariant rather than gauge invariant, we cannot use it to define a two-form current as above. In the linearized case, this interpretation become available again. 

\subsection{\texorpdfstring{$p$}{p}-Form fields}
\label{sec:p-form}

We may also investigate $p$-form fields to find the Noether charge associated with their gauge symmetry. In $d$ dimensions, a $p$-form $A^{(p)}$ has a $(p+1)$-form field strength \[F^{(p+1)} = dA^{(p)}.\] There exists a gauge transformation $\delta_{\Lambda}A^{(p)} = d\Lambda^{(p-1)}$ under which the field strength is invariant. Notice that the gauge parameter itself has a gauge transformation $\delta\Lambda^{(p-1)} = d\Lambda^{(p-2)}$ and so on. Accounting for these additional redundancies, there are $\binom{d-1}{p}$ independent components---off-shell degrees of freedom---in a $p$-form field $A^{(p)}$. An action for the free $p$-form field is given by
\[S = -\frac12\int_{M} \star F^{(p+1)}\wedge F^{(p+1)}.\]
This action is obviously invariant under the gauge variation of the field $A^{(p)}$ and we may use a localized transformation and Noether's second theorem to derive
\[\delta_{R,\Lambda}S = - (-1)^{d-p-1} \int_{\del R} \star F^{(p+1)}\wedge d\Lambda^{(p-1)} -(-1)^{d-p} d \star F^{(p+1)}\wedge \Lambda^{(p-1)}\label{eq:varipform}\]
The statement of Noether's second theorem is $d^{2}\star F^{(p+1)} = 0$, which is a more general form of the statement for Maxwell theory. It easy to see that \eqref{eq:varipform} is a total derivative once again and thus
\[\delta_{R,\Lambda}S = - \int_{\del R} d[\star F^{(p+1)}\wedge\Lambda^{(p-1)}].\] The two-form $k$ is therefore given by \[k = \star (\star F^{(p+1)}\wedge\Lambda^{(p-1)}) = \frac12 \Lambda^{\mu_{1}\cdots\mu_{(p-1)}}F_{\mu_{1}\cdots\mu_{(p-1)}\rho\sigma}dx^{\rho}\wedge dx^{\sigma}.\label{eq:pformtwoform}\]
As for Maxwell theory, where $k$ a special case of \eqref{eq:pformtwoform}, fields coupled to the $p$ form need to obey the Abelian gauge symmetry of $A^{(p)}$ if the theory is to stay consistent. The two-form $k$ is therefore unaffected by other additional fields, however, the associated current $j$ is once again altered by additional terms. Additionally, the commutator of charges is Abelian again.

As we discussed before in Sec.~\ref{sec:ward-identities}, there are nontrivial Ward identities whenever there are residual gauge transformations yielding finite contributions from the variation of the action. To the author's knowledge, the soft behavior of $p$-forms has only been studied in the form of the Kalb--Ramond field in the literature \cite{DiVecchia:2015oba}, where it has been shown that there is no leading soft factor, but a type of subleading soft factor. From a purely technical point of view, it is the antisymmetry of the Kalb--Ramond field's polarization tensor which excludes a leading soft factor. 

Assuming a Lorenz-type gauge condition, $\star\ d\star B^{(2)} = 0$, the residual gauge parameter $\Lambda^{(1)}$ satisfies the Proca equation \cite{PhysRevD.9.2273} \[\star d \star (d\Lambda^{(1)}) = 0\] which clearly has nontrivial solutions. Thus one would expect that there are nontrivial charges from a subsector of the gauge symmetry of the Kalb-Ramond field. At this stage, we have made no attempt to investigate these residual gauge symmetries of the Kalb-Ramond field in detail nor to connect them with the soft behavior of the Kalb-Ramond field. In general, the existence of interesting charges for $p$-forms necessarily depends on the number of dimensions $d$ and the rank $p$. The two-form \eqref{eq:pformtwoform} can also be used to investigate 10D supergravity where large gauge transformations of Ramond-Ramond fields may create nontrivial Ward identities.

\subsection{Gravity}
\label{sec:spin-2-graviton}

Let us turn now to gravity and change the focus slightly. While many of the results found in gauge theories hold, we would like to investigate some effects specific to gravity at this point. The reader may have noticed that we omitted the Rarita--Schwinger field which also has a gauge symmetry. We shall cover supergravity and this fermionic field in particular in a subsequent publication. We should note that the historical context of~\cite{Noether1918} was the question of how to define energy in general relativity~\cite{Kosmann}.

In this section we investigate the Noether two-form that we receive from Noether's second theorem when used on the diffeomorphism symmetry of a linear metric perturbation in Einstein gravity. Interestingly, the two-form Noether charge for diffeomorphisms is well known in the literature \cite{Wald:1993nt,Iyer:1994ys,Wald:1999wa,Barnich:2001jy} and has been used for a wide variety of problems.\footnote{See also e.g., \cite{Nazaroglu:2011zi,Azeyanagi:2014sna}.} It is connected to the entropy of a black hole \cite{Wald:1993nt} and has been connected with information theoretic measures \cite{Lashkari:2015hha}. It has been used to derive the soft graviton theorem \cite{He:2014laa,Strominger:2014pwa,Strominger:2013jfa,Kapec:2015vwa}. The fact that it can be derived with the help of Noether's second theorem, however, has not been sufficiently emphasized in the literature.

Starting with the Einstein--Hilbert action with Gibbons--Hawking--York boundary term
\begin{equation}
S = S_{EH} + S_{GHY}\qquad
S_{EH} = \frac{1}{2\kappa^2}\int_M d^d x\sqrt{-g} R\qquad
S_{GHY} = \frac{1}{\kappa^2}\int_{\partial M} d^{d-1} x\sqrt{-\gamma} K,
\end{equation}
where $\gamma$ is the metric on $\partial M$, one may follow the procedure outlined in Sec.~\ref{sec:noeth-second-theor} and then write Ward identities as in~\ref{sec:ward-identities}. There is an additional complication, beyond the treatment there, in that the action depends on two derivatives of the metric. This leads to a $\nabla_\mu\nabla_\mu \rho$ term in~\eqref{eq:deformed-deltaS}. One might consider avoiding this issue by  switching to a first-order formulation; however, the extra term affects neither the Ward identity nor the two-form $k$, assuming the boundary term is consistent with the variational principle. One finds the two-form in~\cite{Iyer:1994ys}:
\begin{equation}
k^{\mu\nu} = \frac{1}{2\kappa^2}(\nabla^\mu\xi^\nu - \nabla^\nu\xi^\mu).\label{eq:wy2form}
\end{equation}
In order to avoid several complications, it is convenient to work with linearized gravity instead of the full nonlinear theory. Indeed, we don't know how to work with the path integral of the nonlinear theory, in any case.

Assuming a perturbation $h_{\mu\nu} = g_{\mu\nu} - \overline{g}_{\mu\nu}$ where we use $\overline{g}_{\mu\nu}$ as a classical vacuum-like background and $g_{\mu\nu}$ a metric that satisfies the nonlinear equations of motion of gravity with some matter energy-momentum tensor $T^{\mu\nu}$
\[
R_{\mu\nu}-\frac12 g_{\mu\nu}R = \kappa^{2}T_{\mu\nu},
\] 
the equations of motion for the perturbation are given by 
\[
E^{\mu\nu} = \left(\overline\nabla{}^{\mu}\overline\nabla{}^{\nu}h + \overline\nabla{}^{\lambda}\overline\nabla_{\lambda}{}h^{\mu\nu} - 2\overline\nabla_{\lambda}\overline\nabla{}^{(\mu}h^{\nu)\lambda}-\overline{g}{}^{\mu\nu}(\overline\nabla{}^{\lambda}\overline\nabla_{\lambda}h - \overline\nabla_{\lambda}\overline\nabla_{\rho}h^{\lambda\rho})\right).
\] 
More generally, a linearized equation of motion for gravity coupled to matter is given by
\[
E_{\mu\nu} = \frac{1}{2\kappa^{2}}\Big(R^{L}_{\mu\nu}-\frac12\overline{g}_{\mu\nu}R^{L}\Big)=(T_{\mu\nu} + t_{\mu\nu}) = \mathcal{T}_{\mu\nu}
\] 
where superscript $L$ indicates linearized quantities. We also included $t_{\mu\nu}$, which are the higher order terms of the expansion of the Einstein tensor.

Using Noether's second theorem \[2 E^{\mu\nu}\overline\nabla_{\mu}\xi_{\nu} = \overline\nabla_{\mu}(2E^{\mu\nu}\xi_{\nu}) = \overline\nabla_{\mu}S^{\mu}(E,\xi)\] and the current $j^{\mu}$ (which can be derived from a rather tedious calculation), we find the two-form
\[k^{\mu\nu} = \frac{1}{2\kappa^{2}}\left(\xi_{\rho}\overline\nabla_{\sigma}H^{\rho\sigma\nu\mu}+\frac12H^{\rho\sigma\nu\mu}\overline\nabla_{\rho}\xi_{\sigma}\right)\] with \[H^{\mu\alpha\nu\beta} = \frac12\Big( \overline{h}{}^{\alpha\nu}\overline{g}^{\mu\beta} + \overline{h}{}^{\mu\beta}\overline{g}^{\alpha\nu} - \overline{h}{}^{\alpha\beta}\overline{g}^{\mu\nu} - \overline{h}{}^{\mu\nu}\overline{g}^{\alpha\beta}\Big)\]
and $\overline{h}{}^{\mu\nu}$ is the trace reversed metric. We also could have perturbed eq.~\eqref{eq:wy2form} for this expression. The same expression may be found in \cite{Barnich:2001jy} and was obtained originally in \cite{abbott1982stability}. We excluded the energy-momentum tensor $\mathcal{T}_{\mu\nu}$ from the calculation because it will not appear in $k^{\mu\nu}_{\xi}$. The quantity $H^{\mu\nu\alpha\beta}$ is known as the \emph{superpotential} in the literature and it enjoys the same symmetries as the Riemann tensor $R^{\mu\nu\alpha\beta}$. Again, the leading order of this two-form is \emph{linear} in the field $\overline{h}{}^{\mu\nu}$, suggesting that the linear perturbation to the metric field behaves like a Goldstone boson in the sense explained above. One could have included a cosmological constant in the action; the two-form $k$ would not have been changed.

Let us mention that for the case of a Killing vector of the background \[\overline{\nabla}_{\mu}\xi^{i}_{\nu}+\overline{\nabla}_{\nu}\xi^{i}_{\mu}=0\] where $i$ enumerates the number of Killing vector fields, one finds that $k_{\xi}^{\mu\nu}$ is the quantity traditionally used to calculate conserved charges at spacial infinity $i^{0}$ (let $k$ be a tensor density by multiplying with $\sqrt{-\overline{g}}$)
\[M^{i} = \int_{\sigma^{0}} k^{i}.\] Crucially, then, 
\[j^{\mu} = \overline{\nabla}_{\nu}k^{\mu\nu} = \mathcal{T}^{\mu\nu}\xi_{\nu}.\] Note that we do not need to prove that the conserved quantity associated with the Killing vectors are codimension two surface integrals because it is manifestly the divergence of a two-form by Noether's second theorem. 

There is a general path integral formalism available for metric perturbations of Einstein gravity \cite{Bern:1990bh} which we use here to make contact with the soft theorem \cite{Strominger:2013jfa} \emph{at tree level}. For the large diffeomorphisms in asymptotically flat space, or equivalently asymptotic Killing vectors satisfying $\overline{\nabla}_{(\mu}\xi_{\nu)}=O(\nicefrac{1}{r})$ at $\scri$, the formalism developed in Sec.~\ref{sec:ward-identities} together with \cite{Strominger:2013jfa} implies choosing $\rho$ such that we are integrating the two-form over the surface $\sigma$ (a sphere) at $\scri^{+}_{-}$ \[Q[\xi] = \int_{\sigma}\star k_{\xi}\] this integral can then be turned into an integral over all of\footnote{Note, that we are assuming good behavior when approaching timelike infinity along $\scri$ in the spirit of \cite{Strominger:2013jfa}.} $\scri$ where
\begin{align}
\label{eq:starthere}
Q(\xi) &= \int_{\scri}\star (\star\ d\star k_{\xi})\\ &= \frac{1}{2\kappa^{2}}\int_{\scri} dS_{\nu} \Big(2\kappa^{2}\mathcal{T}^{\mu\rho}\xi_{\rho} + \overline{\nabla}_{\mu}(H^{\nu\rho\sigma\mu} + \frac12 H^{\rho\sigma\nu\mu})\overline{\nabla}_{\rho}\xi_{\sigma} + \frac12 H^{\nu\mu\rho\sigma}(\overline{\nabla}_{\mu}\overline{\nabla}_{\rho}\xi_{\sigma} - \overline{R}^{\kappa}{}_{\mu\rho\sigma}\xi_{\kappa})\Big)\notag
\end{align}

We used once again $k_{\xi} = \sqrt{-\overline{g}}\ k$. In the second line, we pulled out the current $S^{\mu}$ using the linearized Einstein equations in the form (see e.g. \cite{abbott1982stability})
\[2\kappa^{2}\mathcal{T}^{\mu\nu} = \overline{\nabla}_{\alpha}\overline{\nabla}_{\beta}H^{\mu\nu\alpha\beta}+\frac12 \overline{R}^{\nu}_{\rho\alpha\beta}H^{\mu\rho\alpha\beta}.\] Clearly, this result is more involved than the case when $\xi$ is a Killing vector. In the Killing vector case, one can use that $\overline{\nabla}_{\alpha}\overline{\nabla}_{\beta}\xi_{\rho} = \overline{R}{}^{\lambda}_{\alpha\beta\rho}\xi_{\lambda}$ to make the last term vanish. For asymptotically flat space times, this term is still very suppressed at asymptotic infinity since the background Riemann tensor $\overline{R}_{\mu\nu\alpha\beta}$ behaves at least as $O(r^{-3})$. 

We can now use Sec.~\ref{sec:ward-identities} with the path integral for ``quantum'' gravity, insert our result from above and retrieve Weinberg's soft theorem in the case of an asymptotically flat manifold. The procedure is very similar to what was presented in Ssec.~\ref{sec:spin-1-gauge}. One first chooses Bondi coordinates and specializes $\xi$ to BMS supertranslations. For these, the leading term at the boundary is $\xi^{v} = T$, where $T$ is an arbitrary function of the coordinates of the sphere at infinity.

To make contact with \cite{Strominger:2013jfa}, we note that we can use Sec.~\ref{sec:ward-identities} with the path integral for ``quantum'' gravity \cite{Bern:1990bh}, insert our result from above and retrieve Weinberg's soft theorem in the case of an asymptotically flat manifold. The procedure is very similar to what was presented in Ssec.~\ref{sec:spin-1-gauge}. One first chooses Bondi coordinates where
\begin{align}
\overline{g}_{\mu\nu}dx^{\mu}\otimes dx^{\nu} &= -du^{2}-2dudr +2r^{2}\gamma_{z\bar{z}}dzd\bar{z}\\
h_{\mu\nu}dx^{\mu}\otimes dx^{\nu} &= 2\frac{m_{B}}{r}du^{2}-2U_{z}dudz-2U_{\bar{z}}dud\bar{z} + rC_{z z}dz^{2} + rC_{\bar{z}\bar{z}}d\bar{z}^{2}
\end{align}
 and specializes $\xi$ to BMS supertranslations
\[\xi^{\mu}\del_{\mu} = T\del_{u}-\frac{1}{r}(D^{\bar{z}}T\del_{\bar z} + D^{z}T\del_{z}) + D^{z}D_{z}T\del_{r} + o(r^{-1})\] where $T = T(z,\bar z)$.
With these coordinates, the only component of $H^{\mu\nu\kappa\lambda}$ appearing in $Q(\xi)=\delta_{R,\xi}S_{EH}$ is $H^{urur} = - \overline{h}^{uu}\overline{g}^{rr} = -2 \frac{m_{B}}{r}$.
For the BMS transformations, the leading term at the boundary is $\xi^{u} = T$ such that we end up with
\[Q(\xi) = \frac{1}{\kappa^{2}}\int_{\sigma}dzd\bar{z}\ \gamma_{z\bar{z}}Tm_{B}\] which we recognize as the BMS charge defined in \cite{Strominger:2013jfa}. We could also have started from \eqref{eq:starthere} to get to the result integrated over all of $\scri^{\pm}$. As before, we may insert this into the path integral for a set of insertions $\Phi_{i}$ by the method described in \ref{sec:ward-identities}. The resulting identity takes the form of \eqref{eq:ward-for-soft} after rewriting $Q(\xi)$ as an integral over all of $\scri$. We recognize the Ward identity as the leading soft graviton theorem.

\section{Discussion}
\label{sec:discussion}

Noether's second theorem, is an old but somewhat underappreciated tool, which acquires new significance in light of recent developments. By combining the robustness of the path integral formalism with the elegance of Noether's second theorem we find that writing down Ward identities for residual gauge symmetries becomes essentially automatic. We have focused on gauge symmetries for fields that obey bosonic statistics, but we will treat the case of the Rarita--Schwinger field as the gravitino in supergravity in an upcoming publication. There a host of new and interesting effects appears. Much like in the case of the BMS symmetry, which contains Poincar\'e transformations, we find that the enhanced symmetry contains the usual supersymmetry charges of the supergravity background. The resulting Ward identities form the soft theorem of the gravitino. 

Our approach sheds some light on interesting connections that emerge from the literature. Especially important is the interpretation of some residual gauge symmetries as a shift symmetry in the spirit of Goldstone's theorem, as well as the connection between Strominger et al's highly influential soft graviton theorem as a consequence of BMS symmetry and Wald's important discovery that the Noether two-form integrated over the bifurcation two-sphere is the entropy of the black hole.

Additionally, the current investigation opens the path to the study of anomalies for residual gauge symmetries. By anomaly, we mean actual one-loop effects connected to the path integral measure as opposed to classical effects like shifts of the boundary conditions of the path integral, which lead to classical central charges. We know that the subleading soft theorem in gravity as well as the leading and subleading soft theorem in YM theory get corrected at the first loop level from the study of scattering amplitudes. We conclude that some of the residual gauge symmetries cannot survive quantization. This is not a problem, since they are not traditional gauge symmetries; by definition they are not gauge fixed out of the path integral measure. An anomaly in these symmetries, thus, does not signal the breakdown of the theory. 

More speculatively, we alluded several times to the possibility of slicing open the path integral and writing Ward identities for subregions of the total spacetime. It is desirable to explore how that works in detail, and its physical consequences. One may also consider transformations involving the Fadeev--Popov ghosts, which ultimately should lead to statements in the BRST and BV formalisms.

\paragraph{Acknowledgments}
\label{sec:acknowledgment}

BUWS would like to thank Andrew Strominger for useful discussions as well as the Department of Physics at Brown U.\ for continued hospitality. BUWS is supported by the Cheng Yu-Tung fund, the Center for Mathematical Sciences and Applications at Harvard University, and NSF grant 1205550. SGA is grateful for correspondence with Nemani Suryanarayana, and for discussions with Antal Jevicki and Samir Mathur. SGA is supported by US DOE grant DE-SC0010010. The authors are thankful to Matteo Rosso and Bruno Le Floch for useful comments.

\newpage
\printbibliography
\end{document}